\pdfoutput=1
\documentclass[aps,prl,reprint,superscriptaddress,showkeys,amsmath,amssymb]{revtex4-2}
\usepackage{bm}
\usepackage{mathtools}
\usepackage{graphicx}
\usepackage{hyperref}
\newcommand{\vct}[1]{\boldsymbol{#1}}
\newcommand{\dd}{\mathrm{d}}

\begin{document}

\title{Turbulence at the Correlation Scale as the Driver of Magnetic Reconnection: In Situ Observations and Kinetic Simulations}

\author{M. B. Khan}
\email{mbilal@udel.edu}
\affiliation{Department of Physics and Astronomy, University of Delaware, Newark, DE 19716-2570, USA}
\author{M. A. Shay}
\affiliation{Department of Physics and Astronomy, University of Delaware, Newark, DE 19716-2570, USA}
\author{R. Bandyopadhyay}
\affiliation{Department of Physics and Astronomy, University of Delaware, Newark, DE 19716-2570, USA}
\author{S. Adhikari}
\affiliation{Department of Physics and Astronomy, University of Delaware, Newark, DE 19716-2570, USA}
\author{W. H. Matthaeus}
\affiliation{Department of Physics and Astronomy, University of Delaware, Newark, DE 19716-2570, USA}
\author{P. A. Cassak}
\affiliation{Department of Physics and Astronomy, Clemson University, Clemson, SC 29634, USA}
\author{J. E. Stawarz}
\affiliation{Department of Mathematics, Physics and Electrical Engineering, Northumbria University, Newcastle Upon Tyne NE1 8ST, UK}
\author{S. Roy}
\affiliation{Space Research Institute, Austrian Academy of Sciences, Graz, Austria}
\author{S. Oughton}
\affiliation{Department of Mathematics, University of Waikato, Hamilton 3240, New Zealand}
\author{C. C. Haggerty}
\affiliation{Department of Physics and Astronomy, The University of Iowa, Iowa City, IA 52242, USA}
\author{Y. Yang}
\affiliation{Department of Physics and Astronomy, University of Delaware, Newark, DE 19716-2570, USA}
\author{F. Pecora}
\affiliation{Department of Physics and Astronomy, University of Delaware, Newark, DE 19716-2570, USA}
\author{J. M. TenBarge}
\affiliation{Department of Astrophysical Sciences, Princeton University, Princeton, NJ 08544, USA}

\begin{abstract}
We study the statistical properties of magnetic reconnection using high-resolution \emph{in situ} observations from the Magnetospheric Multiscale (MMS) Mission in the terrestrial magnetosheath and kinetic particle-in-cell (PIC) simulations. For the first time, using \emph{in situ} observations, we find that the statistics of upstream reconnecting fields in a macroscopic turbulent plasma system are closely linked with the amplitude of turbulence at the correlation scales. These results are further confirmed with fully kinetic plasma turbulence simulations.
Consequently, the average magnetic reconnection rates and associated heating rates, determined through the upstream fields,
could become orders of magnitude higher than those predicted using the turbulence amplitude at the dissipation scale. We then use a simplified model of the heating of reconnection exhausts to estimate the ratio of this reconnection heating
rate
to the overall energy cascade rate due to turbulence. Estimates from the PIC simulations and the MMS observations yield a ratio on the order of $1/5$,
indicating that
direct reconnection heating is playing a substantial role in dissipating the turbulent fluctuations.

\end{abstract}

\maketitle

Magnetic reconnection is a fundamental plasma process responsible for dissipating magnetic energy in a wide range of space and astrophysical environments, such as the solar corona, solar wind, and terrestrial magnetosphere \cite{Bandyopadhyay2021EnergyReconnection, ShayEA18, OsmanEA11-heat}. In these systems, magnetic reconnection contributes to plasma heating and also drives non-thermal particle acceleration, leading to the formation of suprathermal tails in particle velocity distributions. It also plays a
crucial role in driving explosive events such as solar flares, coronal mass ejections, and geomagnetic substorms. In laboratory settings, it can critically impact confinement and stability in plasma fusion devices, where reconnection events could lead to energy losses and loss of confinement.

Besides exhibiting reconnection, the aforementioned plasma systems -- whether astrophysical, space-based, or terrestrial -- have an intrinsic tendency to develop into a seemingly random and chaotic state known as turbulence~\cite{Bruno2005ThePhysics, MatthaeusEA21, MattGold82a}. A key question in these turbulent systems is how energy injected at large scales cascades to smaller scales and ultimately heats the plasma. While turbulence facilitates this cascade of energy across scales, the specific mechanisms responsible for irreversible heating, particularly in weakly collisional plasmas, are not fully understood.

A number of numerical studies~\cite{Servidio2010StatisticsTurbulence, Servidio2011MagneticTurbulence, Wan2012IntermittentTurbulence, Zhdankin2013StatisticalTurbulence, Dong2022Reconnection-drivenTurbulence} and observations of the solar wind and terrestrial magnetosphere~\citep{RetinoEA07, OsmanEA11-heat, Phan2018ElectronMagnetosheath, StawarzEA22} have shown that reconnection occurs frequently in the intermittent current sheets
present in
turbulence and may be heating the plasma~\cite{PezziEA21}. However, the relative role that magnetic reconnection plays in the damping of turbulent fluctuations remains poorly understood, as it ostensibly requires knowledge of statistics of magnetic fields upstream of reconnection sites, which in turn requires sampling almost all of the reconnection events in the global system, which is not easily accessible through spacecraft observations.
However, if there is a link between
the statistics of upstream reconnection magnetic fields and the global statistics of turbulent fields (the latter being readily accessible through spacecraft observations), we may employ this linkage to quantify the role of magnetic reconnection in turbulent heating.

A recent study shed some light on the intricate connection between the statistics of reconnecting fields and the global turbulent fields using single fluid magnetohydrodynamic (MHD) simulations~\cite{khan2025does}. The study demonstrated that in a large turbulent system the statistics of upstream reconnecting fields are strongly correlated with the statistics of the turbulent magnetic field at the correlation scale of the global system---reinforcing the earlier idea that turbulent dissipation is essentially governed by the large-scale eddies
 \cite[e.g.,][]{BatchelorTHT,Frischbook}.
 This connection, however, has not yet been investigated for low or weakly collisional systems, using either simulations or \emph{in situ} observational data. The current study addresses this gap by employing both \emph{in situ} observational data, from the Magnetospheric MultiScale (MMS) mission in the terrestrial magnetosheath, and fully kinetic particle-in-cell (PIC) simulations. We find that the control of reconnection magnetic fields by correlation scale turbulence also occurs in low collisionality plasmas, establishing the robust nature of this result across a broad range of plasma conditions.

 To quantify the role of magnetic reconnection in the overall heating of the turbulent plasma, we examine
 ``\emph{direct} reconnection heating,''
 which is defined as the heating of plasma as it is processed by reconnection and accelerated into the bulk outflow. We apply our primary result (the reconnection magnetic fields are controlled by the correlation scale) to a simple model of direct reconnection heating and calculate the expected heating rates for  both the PIC simulations and the MMS observations. We find that the direct reconnection heating accounts for approximately 1/5 of the energy cascade rate. Clearly, reconnection heating is likely playing an important role in dissipating the turbulent fluctuations.

    \section*{PIC Simulations}
    \label{sec:PIC}

To study the statistics of magnetic reconnection, we use a 2.5D fully kinetic PIC simulation of decaying plasma turbulence performed using the P3D code \cite{ZeilerJGR02, ParasharEA18}. In this simulation, magnetic fields are normalized to a reference value $B_0$ and the density to a reference value $n_0$. The length and time scales are normalized to the ion inertial length $d_{i} = \sqrt{m_i/(\mu_0\,n_0\,e^2)}$ and inverse ion cyclotron frequency $\omega_{ci}^{-1} = m_i/(e\,B_0)$, respectively.
Speeds are normalized to the ion Alfv\'en speed $c_{A0} = B_0/ \sqrt{\mu_0\,m_i n_0} $. The ion to electron mass ratio is
  $ m_i / m_e= 25$,
and the ratio of the speed of light to the reference Alfv\'en speed is $c/c_{A0} = \omega_{pi}/\omega_{ci} =15$. The simulation was performed in a periodic square domain of length $L_x = L_y = 149.6 \, d_i$, with $4096^2$ grid points and 3200 particles per cell (ppc or ppg) for each species.

The system is initialized with a uniform density $n = 1.0\,n_0$ and a uniform guide field $b_{z0} = 1.0\,B_0$, with uniform ion and electron temperatures $T_{i} = T_{e} = 0.3$. These yield ion and electron plasma betas, $\beta_i = \beta_e = 0.6$, based on the guide field $b_{z0}$. The simulation is initialized with MHD-like Alfv\'enic initial conditions, using gaussian fluctuations with random phases for the in-plane
 ($\hat{\vct{x}}$ and $\hat{\vct{y}}$ components of)
magnetic and velocity fields. The electrons carry the out-of-plane current $j_z$ to generate the magnetic field fluctuations.
The initially excited Fourier modes have wave vectors $\bm{k}$ in the annulus $ 2 \le | \bm{k} | \le 8 $. Initially, the total fluctuation energy is $0.16\,m_i\,c_{A0}^2$ and is equipartitioned between the magnetic and kinetic energy ($\delta b_\text{rms} = 0.4$ and $ \delta u_\text{rms} = 0.4$).   The system evolves without any external forcing, and we analyze it at
 a
time when the out-of-plane current density is near its peak value:  $t=50.5 \, \omega_{ci}^{-1}$. At this time, the turbulence is well developed and energy is well distributed across a broad range of length scales.

To study the reconnection activity, we first identify the X-points as
the
saddle points in the out-of-plane component of magnetic potential. Due to small-scale noise in PIC simulations, large numbers of spurious X-points are created. To address this issue, Haggerty et al.\
  \cite{Haggerty2017ExploringTurbulence}
performed a detailed analysis of X-point identification methods that may be used with kinetic PIC simulation data.
They concluded that
Fourier space filtering at the Debye scale was the most effective of the techniques examined for removing the spurious X-points while still retaining the physical X-points.
Here we employ this same methodology, applying a
(sharp)
Fourier space filter at the Debye scale to the out-of-plane magnetic potential and then analyzing its topographical properties using the Hessian matrix technique~\cite{Servidio2010StatisticsTurbulence}. At the time of analysis considered herein, this method yields a total of 274 physical X-points. Eight of these were removed from our analysis, as they showed negligibly weak reconnection activity, leaving a total of 266 reconnection events for use in the present study.

At each X-point, we first determine the local thickness of the diffusion region using a hybrid method described in detail previously~\cite{khan2025does}. This method estimates the inflow width of the diffusion region by examining $j_z(s)$, i.e., a cut of $j_z$ along the inflow direction $s$. The distance from the X-point to the edge of the diffusion region is the smaller of (1) the location where $j_z = 0$ and (2) the width determined from a curve fit of $j_z(s)$. Note that because some asymmetric reconnection occurs~\cite{CassakShay07}, we calculate the width along both inflow directions independently. These widths are then used to determine the upstream magnetic fields, $\vct{b}_{\text{up}_1}$ and $\vct{b}_{\text{up}_2}$, for each X-line.

To study the statistics of reconnecting fields, we first compute the reconnection increments $ \Delta b_\text{rec} $ at each of our identified reconnection sites. Following a similar approach as in Ref.~\cite{khan2025does}, we define the reconnection increment as
 $ \Delta b_\text{rec}
 = |\vct{b}_{\text{up}_1}  -  \vct{b}_{\text{up}_2}|
 = |\vct{b}_{\text{up}_1}| + |\vct{b}_{\text{up}_2}| $. Examining the probability distribution functions (PDFs) of these reconnection increments reveals the statistics of the magnetic reconnection occurring in the turbulence.

 The key question is how the PDFs of reconnection increments compare with similar statistical properties of the turbulence in general. Noting that the reconnection increments are inherently transverse, we calculate the global transverse turbulence increments $\Delta \vct{b_T}(\vct{x},\vct{r}) = (\vct{b}(\vct{x} + \vct{r}) - \vct{b}(\vct{x}))\cdot\vct{\hat{n}},$ where $\vct{x}$ is the position vector, $\vct{r}$ is the lag vector, and $\vct{\hat{n}}$ is a unit vector perpendicular to $\vct{r}$. We then take the absolute value of these increments, denoting them as $\Delta b_T(\vct{r})$.
 Here, ``global'' means the quantities are calculated across the whole domain (rather than only at the reconnection sites).

A major conclusion of the previous MHD study~\cite{khan2025does} was that the turbulence properties at the correlation scale
(rather than at the scale of the
 thickness of reconnection diffusion regions)
control the statistics of the reconnection. To see if this conclusion extends to low collisionality systems, we compare PDFs of $ \Delta b_\text{rec}/2 $ with those of
the global turbulence increments,
  $ \Delta b_T(r) $,
at two different lags $r$:
(i) the average thickness of the diffusion region $ r= \lambda_\text{rec}$, and
(ii) the correlation scale $ r = \lambda_c$.
We compute the latter by integrating the
 magnetic field
correlation function,
 $ \lambda_c = \int_0^{L_x/2}C(r) \,\dd{r} $
where
 $ C(r) =  \langle
   \vct{b}(\vct{x}+\vct{r}) \cdot  \vct{b}(\vct{x})
   \rangle /
 \langle \vct{b}(\vct{x}) \cdot \vct{b}(\vct{x})
 \rangle $ and $\langle\bullet\rangle$ denotes an average over the simulation box. In our simulations, we find, $\lambda_\text{rec}\approx 1\,d_i$ and $\lambda_c\approx3.5\,d_i$.
\begin{figure}
    \centering
\includegraphics[width=\linewidth]{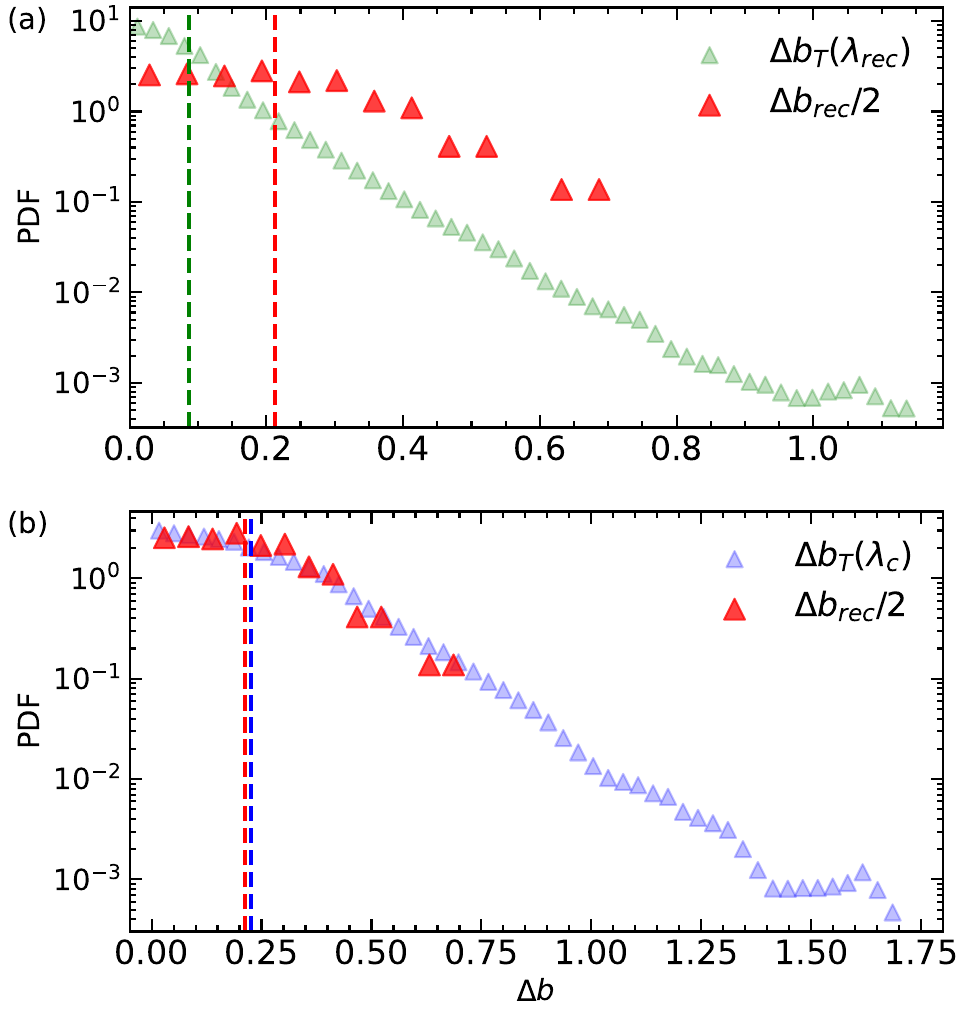}
   \caption{PDFs from the PIC simulation data of reconnection and turbulence increments. Reconnection increments, $\Delta b_\text{rec} $, are computed at the X-points, and shown on both panels (red triangles). Increments for the (global) turbulent magnetic field, $ \Delta b_T(r) $ are computed at two distinct scales: (a) the average transverse reconnection scale $\lambda_\text{rec}$, and (b) the turbulence correlation scale $\lambda_c$.  See main text for full definitions.  Vertical dashed lines indicate mean values of the distributions.
   Note the different horizontal ranges in (a) and (b).}
    \label{fig:PIC}
\end{figure}

\begin{figure}
    \centering
\includegraphics[width=\linewidth]{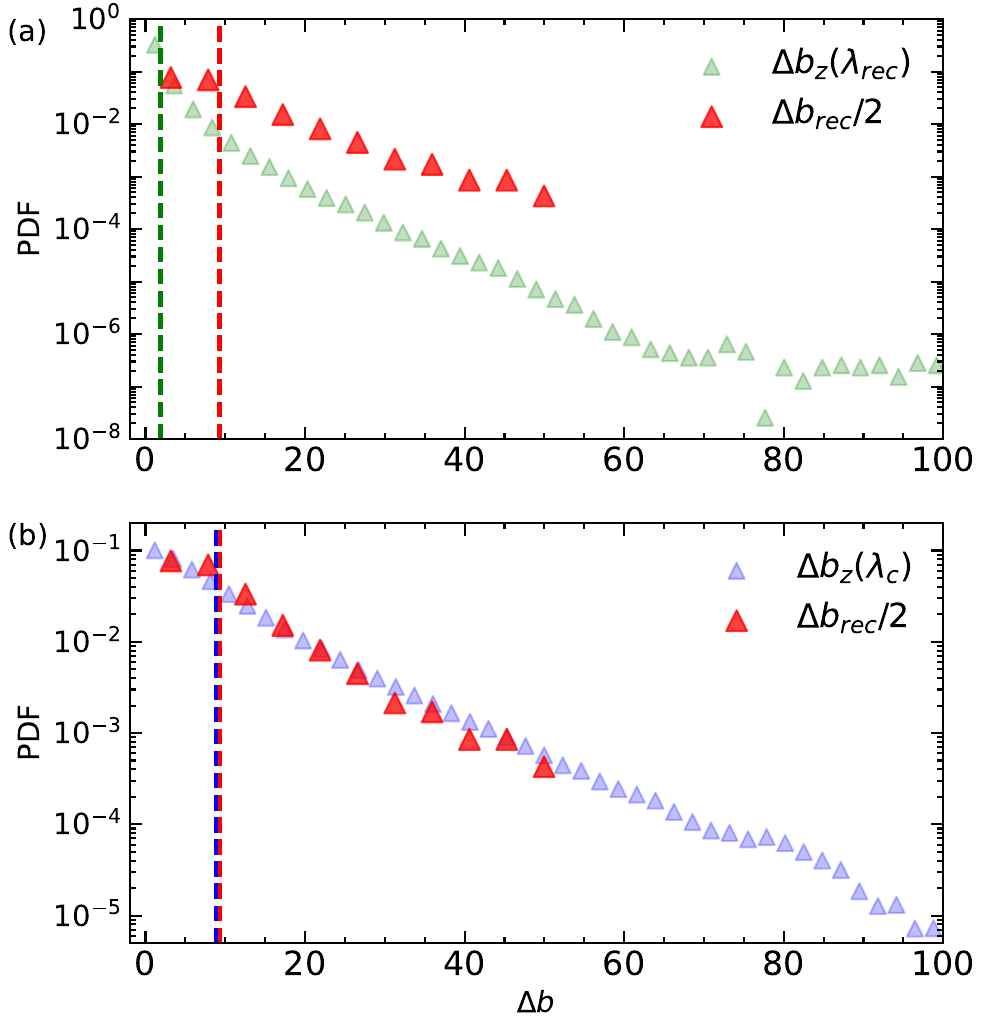}
    \caption{
    Comparison of reconnection and turbulence increments using MMS observations of the turbulent magnetosheath, in the same format as Fig.~\ref{fig:PIC}. The quasi-transverse turbulence increments are calculated using $b_z$ at two different lags in 60 turbulence intervals: the average reconnection region width ${\lambda}_\text{rec}$ and the (individual)
    correlation scales $\lambda_c$. The reconnection increments are calculated from 256 reconnection events
      identified in the 60
    turbulence intervals. }
    \label{fig:MMS}
\end{figure}

Fig.~\ref{fig:PIC}a shows a comparison between the PDF of transverse turbulence increments and the reconnection increments for our PIC simulation. The turbulence increments are evaluated at the average transverse reconnection scale (i.e., the average diffusion region thickness) in the system.
The comparison clearly demonstrates the dissimilarity between the two distributions and their corresponding mean values.
Following the same line of argument presented in a recent
 MHD
study
  \cite{khan2025does},
we perform a
second such comparison
but this time with the (global) transverse turbulence increments evaluated at the \emph{correlation} scale
 (Fig.~\ref{fig:PIC}b).
Perhaps surprisingly, the global field statistics at the correlation scale show a strong correlation with the statistics of the reconnecting fields.
Their mean values and their PDFs are in agreement.
These results from fully kinetic simulations---which retain the physics of both ions and electrons---are strikingly similar to those obtained from a study performed using incompressible MHD
(i.e., single fluid)
simulations~\cite{khan2025does}.

An immediate extension of these results is to investigate their validity using \emph{in situ} observations of a kinetic plasma turbulence system, such as the terrestrial magnetosheath.  We explore these ideas in the following discussion.

    \section*{MMS Observations}
    \label{sec:MMS}

The MMS mission provides high-cadence observational data for plasma processes in the near-Earth system \cite{BurchEA16}.
The small
separations between the four spacecraft allow us to resolve key scales associated with kinetic plasma turbulence with unprecedented accuracy. In this study, we use a statistical sample of 256 reconnection events identified in a previous publication using MMS observations of shocked solar wind in the terrestrial magnetosheath~\cite{StawarzEA22}. These events were found within 60 data intervals observed between October 2015 and May 2018.
In Geocentric Solar Ecliptic (GSE) coordinates, the locations of the intervals are broadly distributed around the near-equatorial magnetosheath, ranging from the sub-solar point up to $ 132^\circ $
into the flanks. The data for magnetic field and plasma velocities are obtained from the fluxgate magnetometer (FGM) \cite{Russell2016} and the fast plasma instrument (FPI) \cite{PollockEA16}, respectively, onboard the MMS spacecraft. In burst mode, these instruments provide the magnetic field, ion, and electron measurements at cadences of
128\,Hz, 150\,ms, and 30\,ms, respectively. Further details can be found in Ref.~\cite{StawarzEA22} and its supplementary material.

To study the statistics of reconnection magnetic fields, we first compute the reconnection increments
at each of the identified reconnection events. Following Ref.~\cite{StawarzEA22}, a local LMN coordinate system is established for each event using the hybrid minimum variance analysis (HMVA) technique~\cite{GoslingPhan13}, where $\vct{L}$ is the outflow direction, $\vct{N}$ is the inflow direction, and $\vct{M}$ is perpendicular to the reconnection plane, completing the right-handed triad. We sample the reconnecting field components, $\vct{b}_{\text{L}_1}$ and $\vct{b}_{\text{L}_2}$, at the start and end times of each reconnection event. These times, as well as the LMN directions, are provided in the supplementary material of Ref.~\cite{StawarzEA22}. The reconnection increments are then given by $\Delta b_\text{rec} = | \vct{b}_{\text{L}_2} - \vct{b}_{\text{L}_1} |$. In most cases, all four spacecraft sample both inflow regions of each reconnection event. Thus, for each event there are multiple measurements of inflowing plasma properties and magnetic fields. For these 256 reconnection events, data from all four spacecraft yield 1001 reconnection increments.

Additionally, for each of the 60 MMS turbulence intervals, turbulence increments are calculated in the following way. First, the average ion bulk flow velocity $u_0 = | \langle \vct{u} \rangle |$ is calculated
 (these values are available in the supplementary material for \cite{StawarzEA22}).
Then, the relationship between the spatial lag
 ($r$)
and the time lag
 ($t_r$)
is determined using Taylor's hypothesis $r = u_0\,t_r$. The increments for each interval are then given by $\Delta \vct{b} = \vct{b}(t + t_r) - \vct{b}(t)$.

Note that in nearly all of the turbulence intervals, the average bulk flow in the $xy$-plane (in GSE coordinates) is much larger than the flow along $z$.  Therefore, the spatial lags, inferred through Taylor's hypothesis, may be considered effectively limited to the $xy$-plane. Consequently, the $z$-component of the magnetic field increments, being approximately perpendicular to the mean bulk flow, retains a quasi-transverse character~\cite{ChhiberEA18}.
Since we wish to investigate the behavior of transverse increments,
we restrict our MMS analysis to just the $b_z$ increments.

Similar to the PIC analysis,
the turbulence increments are computed at two characteristic scales associated with magnetic reconnection, namely the average thickness of the diffusion region,
    $ \lambda_\text{rec}$,
and the correlation scale, $\lambda_c$. For each of the reconnection events, the spatial thickness of the reconnection region is determined through Taylor's hypothesis using the component of ion bulk flow velocity normal to the reconnecting current sheet, averaged over $10\,d_i$ surrounding the event, and the time duration of the crossing as measured by a single spacecraft (MMS1). These lengths are then averaged to determine $\lambda_\text{rec} \approx 51\,\mathrm{km}$, which is roughly equal to the ion inertial length based on the averaged density of $35\,\mathrm{cm}^{-3}$. The other spacecraft yield similar values for the average thickness of the diffusion region. The correlation scale varies between the different turbulence intervals. Therefore, we use the correlation scale of each individual interval
 (rather than an average $\lambda_c$)
to compute the correlation scale increments.
The collection of increments over all 60 turbulence intervals forms a set used for statistical comparison with the reconnection increments.

In Fig.~\ref{fig:MMS}a, we display the PDF of increments of the turbulent magnetic field at the
(average)
reconnection scale
     ($ \lambda_\text{rec} $)
and the PDF of reconnection increments, $\Delta b_\text{rec}/2$, computed across all of the reconnection sites. The two distributions are clearly distinct and show no indication of being statistically similar.

The situation is quite different for
Fig.~\ref{fig:MMS}b,
which again compares the reconnection increments to turbulence ones, but now the latter are computed at the correlation scale of the turbulence.
These distributions show striking statistical similarity in both their overall shape and mean values, indicating a strong correlation between reconnecting fields and turbulent fields at the correlation scales.
Remarkably, these results closely match those obtained using
(a) the fully kinetic PIC simulations discussed in the previous section, and
(b) the incompressible MHD simulations reported on in an earlier study~\cite{khan2025does}.
The similarity in these results obtained using fluid simulations, PIC simulations, and magnetosheath observations suggests that these findings are not limited to a specific regime of collisionless or collisional plasma turbulence, but may reflect a broader, more general relationship between turbulence and reconnection.

    \section*{Reconnection Heating in Turbulence}
    \label{sec:Heating}

An important question concerns the role that reconnection plays in dissipating the energy of the turbulent fluctuations.
In the last decade, significant progress has been made in understanding the direct heating of plasma in the reconnection exhaust
~\cite{PhanEA14, PhanEA13, ShayEA14, HaggertyEA15, Yamada2014, WangEA16}. Note that this ``direct reconnection heating'' only includes part of the ultimate heating due to reconnection. For example, the compression of plasma at the downstream edge of the exhaust can further heat the plasma ~\cite{Wu&ShayEA12, BarbhuiyaEA22}.

Applying these heating predictions to turbulent reconnection
allows estimation of the direct heating of the plasma due to reconnection. Frameworks for estimating these values have been established~\cite{ShayEA18,StawarzEA22}, although there are significant uncertainties, including the fraction of plasma processed by the reconnection sites and the number of reconnection sites expected.

Fortunately, with the present work we have resolved one of these ambiguities, namely,
how the upstream magnetic fields scale with the turbulence amplitude. For the present purposes, we present here
a simplified model of heating due to reconnection
in turbulent plasmas that provides an informed but approximate
indication of how much direct reconnection heating
may be occurring, based on
both PIC simulations and observations.

We consider a simplified quasi two-dimensional system having square cross-section of side $L$ and extent $L_z$ along the invariant out-of-plane direction. The system has a total of $N$ magnetic flux ropes that are actively participating in the process of reconnection, only through pairwise interactions. Therefore, the total number of X-lines is of the order of $\frac{N}{2}$. Let us consider first a single X-line in our model system.
The magnetic energy per unit time,  $w_\times$, carried into the reconnection region via the Poynting flux is approximated as:
\begin{equation}
    w_\times  = 2\; \left( \frac{E_\text{rec} b_\text{up}}{\mu_0}\right)\ell L_z,
 \label{eq:eps-f}
\end{equation}
where
$ b_\text{up}$ is the reconnecting magnetic field in the inflowing region,
$ E_\text{rec} \approx V_\text{in} b_\text{up} $ is the reconnection electric field,
$ V_\text{in} $ is the speed of the inflowing plasma,
and
$ \ell$ is the full length of the reconnecting region along the outflow direction.
Note that we are assuming a quasi-2D reconnection site with spatial extent along the reconnection electric field (out-of-plane direction) of size $L_z$. The factor of 2 in \eqref{eq:eps-f} is due to the reconnection region drawing in magnetic energy from both magnetic flux ropes.

Next, we make use of the relation
$V_\text{in} = R\,c_{A\text{up}} $,
where $c_{A\text{up}} = b_\text{up} / \sqrt{\mu_0 m_i n} $ is the Alfv\'en speed in the inflow region
and
$ R $ is the aspect ratio of the diffusion region, which typically has a value of around 0.1 for low collisionality systems~\cite{ShayEA99S}.
Note that $R$ is also called the ``normalized reconnection rate.'' We assume a fraction $\alpha$ of this inflowing magnetic energy $w_\times$ heats the inflowing plasma, giving the heating rate $q_\times \equiv \alpha\,w_\times$ of a single X-line:
\begin{equation}
  q_\times = 2\, \alpha\, R\,c_{A\text{up}} \left( \frac{b_\text{up}^2}{ \mu_0} \right) \ell L_z\;.
\label{eq:Q}
\end{equation}

The average heating rate per unit mass, $\epsilon_\text{rec}$, for the entire system is calculated by summing over the reconnection sites and dividing by $m_i\,n\,L^2L_z,$ the total mass in the system. Assuming that the mass density is relatively uniform, we obtain
\begin{equation}
    \epsilon_\text{rec} \approx \sum_{k=1}^{\frac{N}{2}} \left( \frac{2\, \alpha\, R}{L^2} \right) \ell_k\,c_{A\text{up},k}^3\;.
    \label{eq:Q_Tsum}
\end{equation}

The primary difficulty
at this point is performing the sum. Regarding $\ell_k,$ we argue that the average cross-section diameter of a flux rope is $ \approx 2\,\lambda_c$,
where $\lambda_c$ is the correlation scale of the turbulent fluctuations.
Given that a reconnection event involves two flux ropes, a reasonable estimate of the average length of the reconnection region along the outflow direction is $\langle \ell \rangle_\times \approx 2\,\lambda_c$, where $\langle \bullet \rangle_\times $ is an average over all of the X-lines in the system. The number of X-lines in the system is then approximated as $\frac{N}{2} \approx \frac{1}{2}\,\left( \frac{L}{2\,\lambda_c} \right)^2.$

For the present study, $c_{A\text{up},k}$ is directly available from both the observations and the PIC simulations, using $b_{\text{up},k} \approx \Delta b_{\text{rec},k}/2$, allowing us to directly calculate $ \sum_{k=1}^{\frac{N}{2}} \;c_{A\text{up},k}^3 = \frac{N}{2}\,\langle c_{A\text{up}}^3 \rangle_\times$. Substituting for $N$ yields
\begin{equation}
    \epsilon_\text{rec} \approx \frac{1}{2}\,\alpha\,R\;\frac{\langle c_{A\text{up}}^3 \rangle_\times}{\lambda_c}\;.
    \label{eq:reconn-heating}
\end{equation}
This expression is remarkably similar to the von K\'arm\'an
  ``$ u_\text{rms}^3 / \lambda_c $''
heating rate written for a magnetofluid~\cite[e.g.,][]{Hossain96}.

Below we will apply this model
for the direct reconnection heating rate
to the PIC and MMS datasets we have employed herein. These estimates for $ \epsilon_\text{rec} $
will then be compared to the total dissipation
rates for these systems, $ \epsilon_\text{obs}$.

\subsection*{Estimate of reconnection heating in the PIC simulation} \label{sec:PIC-heating}

We first apply the direct reconnection heating  given by \eqref{eq:reconn-heating} to the kinetic PIC simulations. We focus in this study on the ion heating associated with reconnection.
In normalized units, the heating rate per unit mass is given as:
\begin{equation}
    \epsilon_\text{obs} \equiv \frac{\dd}{\dd t} \left( \frac{\sum_{\text{grids}} \frac{1}{2}\, \text{Tr} \left[ \mathbf{P}   \right]  }{\sum_{\text{grids}} n  } \right),
    \label{eq:sim-heating-rate}
\end{equation}
where $\mathbf{P}$ is the pressure tensor and $n$ is the ion number density.
 As
the mean normalized density in the simulation is unity, \eqref{eq:sim-heating-rate} is equivalent to averaging $(1/2)\, \text{Tr} [\mathbf{P}]$ over all grid points.
This calculation yields $\epsilon_\text{obs} \approx 3\times10^{-4}$ during the time of analysis of Fig.~\ref{fig:PIC}.

We then estimate the direct reconnection heating from \eqref{eq:reconn-heating}, noting that its form in simulation normalized units is unchanged.
For each X-line,
$c_{A\text{up}}$
is calculated
 using the data shown in
Fig.~\ref{fig:PIC}, i.e., the magnetic fields used in $c_{A\text{up}}$  are the $\Delta b_\text{rec}/2$; conveniently, $n \approx 1$ throughout the simulation.  The simulation yields $\lambda_c \approx 3.5$ and  $\langle c_{A\text{up}}^3 \rangle \approx 0.024$.
Estimating $\alpha \approx 0.2$, which is a typical value~\cite{PhanEA14},
and $R \approx 0.1$ for fast kinetic reconnection,
\eqref{eq:reconn-heating} yields
 $ \epsilon_\text{rec} \approx 7 \times 10^{-5}.$

An important quantity is $\epsilon_\text{rec}/\epsilon_\text{obs} \approx \frac{1}{4},$ which implies that approximately a quarter of the total ion heating is due to direct reconnection heating.
We
  refer to
this ratio as the ``fractional reconnection heating.''
Clearly, magnetic reconnection is likely playing an important role in facilitating dissipation of turbulent fluctuations.

It is
  also
illustrative to examine the PIC simulation as one would a generic turbulence observation, where the number of X-lines as well as their upstream values are unknown. In that case, $\langle c_{A\text{up}}^3 \rangle$ is not available.
Nonetheless,
it is straightforward to estimate $\langle c_{A\text{up}} \rangle$ in two different ways:
(A) By examining transverse  increments of the turbulent magnetic field at the correlation scale,
or
(B) by directly calculating the rms magnetic field of one of the fluctuating components.
For (A), Fig.~\ref{fig:PIC} reveals that an excellent proxy for $\langle c_{A\text{up}} \rangle$ is the average value of the transverse magnetic increment at the correlation scale, $\Delta b_T (\lambda_c) \approx 0.22$.
For method (B),
the rms value of $b_x$ in the simulation yields $0.27$.
For comparison purposes, the simulation directly yields
  $\langle c_{A\text{up}} \rangle \approx 0.21$.
Substituting $\langle c_{A\text{up}} \rangle^3$ for $\langle c_{A\text{up}}^3 \rangle$ in \eqref{eq:reconn-heating} yields $\epsilon_\text{rec} \approx 3 \times 10^{-5},$ a factor of two smaller than the actual value, but still the right order of magnitude.

\subsection*{Estimate of reconnection heating in the magnetosheath}

The MMS dataset utilized in this manuscript~\cite{StawarzEA22} provides a unique opportunity to estimate the heating rate directly from  observationally measured statistics of reconnection. While this dataset has been used to estimate possible ranges of heating rates previously~\cite{StawarzEA22},  in this study we focus on an average heating rate in line with the analysis leading up to \eqref{eq:reconn-heating}.  A key quantity is the average upstream reconnection magnetic field $\langle b_\text{up} \rangle_\times$. To directly connect with the simulation analysis, we estimate
$\langle c_{A\text{up}}^3 \rangle_\times \approx \langle b_\text{up}^3 \rangle_\times / (\sqrt{\mu_0 m_i \, n_\text{av}})^3$, where $n_\text{av}$ is the average density computed using all of the turbulence intervals. For the correlation scale, we use the average correlation scale over all of the turbulence intervals.

Using
$\langle b_\text{up}^3 \rangle_\times \approx 2.9 \times 10^3\,(\mathrm{nT})^3$,
$n_\text{av} \approx 35 \, \mathrm{cm}^{-3}$,
$\lambda_c \approx 2\times 10^3 \, \mathrm{km}$,
$\alpha \approx 0.2$,
and $R \approx 0.1$
  in \eqref{eq:reconn-heating}
we obtain
\begin{equation}
    \epsilon_\text{rec} \approx 0.7 \times 10^6 \, \mathrm{J} \, \mathrm{kg}^{-1} \, \mathrm{s}^{-1}.
\end{equation}
Studies of magnetosheath turbulence have found cascade rates ranging from about $1 \times 10^6\,\mathrm{J\,kg^{-1}\,s^{-1}}$ up to $20 \times 10^6\,\mathrm{J\,kg^{-1}\,s^{-1}}$~\cite{HadidEA18,BandyopadhyayEA18, PecoraEA23a}. Here we apply the ``lag polyhedral derivative ensemble''~\cite{PecoraEA23a, Pecora2023EAb, PecoraEA24,PecoraEA26} technique to estimate the cascade rate in the 60 MMS turbulence intervals employed in this study, which yields an average value of about $\epsilon_\text{obs} \approx 4 \times 10^6\,\mathrm{J\,kg^{-1}\,s^{-1}}$.
This value is very similar to estimates of cascade rates in the inertial range,
 obtained
using the Taylor hypothesis~\cite{HadidEA18,BandyopadhyayEA18} and assuming isotropy. Note that this average cascade rate excludes turbulence intervals that exhibited a negative cascade rate.

This analysis
 leads to
a fractional reconnection heating of $\epsilon_\text{rec}/\epsilon_\text{obs} \approx 0.2$.
 In other words,
direct reconnection heating
 produces
about 20\% of the total turbulent heating for these MMS reconnection regions, which is similar to
the PIC simulations.

    \section*{Conclusions}
    \label{sec:concl}

The current study aims to explore a fundamental link between the statistical properties of magnetic reconnection and those of turbulence in a macroscopic plasma system. To reveal these connections, we first study fully kinetic PIC simulations of decaying plasma turbulence. In these simulations, at the time of peak turbulence activity, we identified all of the reconnection events using Hessian matrix analysis~\cite{ServidioEA09, Servidio2010StatisticsTurbulence}. We then compared the PDF of reconnection magnetic field increments with the PDF of turbulent magnetic field increments computed at two different scales, namely the correlation scale and dissipation scale. The comparison shows that the reconnecting fields in our system are strongly linked with the turbulent magnetic field at the correlation scale, and not strongly linked with the turbulent field at the dissipation or kinetic scales.

We further explored this connection in the terrestrial magnetosheath plasma, using \emph{in situ} measurements made by  MMS spacecraft. Using a total of 256 reconnection events identified in 60 burst-mode MMS intervals~\cite{StawarzEA22}, we performed a similar analysis to that employed with the PIC simulation data. Comparing the PDFs of reconnection increments and turbulence increments we again find strong agreement between the PDF of reconnection increments and that of turbulence increments calculated at the correlation scale.

Our
results represent a significant departure from previous predictions, in which the reconnection field amplitude is linked to the turbulence amplitude at the (much smaller) dissipation scale~\cite{ShayEA18,LoureiroEA17}.
Note also that the kinetic PIC simulations found heating rates for the ions and electrons individually that were not consistent with a theoretical scaling prediction based on the turbulence amplitude at the ion inertial length~\cite{ShayEA18}.
 Moreover, the consistency between the results obtained from collisional simulations \cite{khan2025does} and from the collisionless simulations and low-collisionality observations
 reported on herein
 further establishes the robustness and universality of this reconnection--turbulence connection.

The present study has wide implications for heating and dissipation in a multitude of astrophysical systems (both collisional and collisionless) where the process of magnetic reconnection occurs as an element of turbulence. Previous models of reconnection rates during turbulence estimated the reconnection magnetic field $b_\text{up}$ using the turbulence amplitude at the reconnection scale~\cite{ShayEA18, LoureiroEA17}. However, the simple model in this paper, \eqref{eq:reconn-heating},
suggests that the dissipation rate due to reconnection is of the same structural form as the turbulence dissipation rate:
    $ c_{A\text{up}}^3 /\lambda_c \sim u_\text{rms}^3 / \lambda_c $.
Using this refined estimate, the predicted reconnection rates and associated heating rates would become orders of magnitude higher than previously expected. Earlier MHD studies have hinted at this connection, both in equipartition turbulence~\cite{khan2025does} and magnetically dominated systems initiated with uniformly sized flux tubes~\cite{ZhouEA18}.
  To the best of our knowledge, however, the present study
 is the first to
 explicitly demonstrate these connections using fully kinetic simulations and, independently, using
\emph{in situ} spacecraft observations.

Following previous studies~\cite{ShayEA18, StawarzEA22}, we
  also
calculate a simple estimate of the rate of  direct reconnection heating, $\epsilon_\text{rec}$,
expected from a turbulent system with a given average upstream Alfv\'en speed and correlation scale.
 By ``direct reconnection heating,'' we mean the heating of the quasi-steady reconnection exhaust plasma,
 which has been the subject of many studies~\cite{ShayEA14,PhanEA13,PhanEA14}. For the PIC simulations and the magnetosheath events, we then provide estimates of the ``fractional reconnection heating,'' $\epsilon_\text{rec}/\epsilon_\text{obs}$, where $\epsilon_\text{obs}$ is the turbulence cascade rate.

For the PIC simulation, we find $\epsilon_\text{rec}/\epsilon_\text{obs} \approx \frac{1}{4}$, and for the MMS observations, the ratio is about $\frac{1}{5},$ meaning that in both cases direct reconnection heating is playing an important role in dissipating the turbulent fluctuations. We emphasize that these ratios may actually be a lower bound on the heating due to reconnection. The heating estimate does not include such effects as the compression of the reconnection exhaust~\cite{BarbhuiyaEA22} in the ``dipolarization front'' or acceleration of particles in secondary magnetic islands~\cite{DrakeEA06}.

An interesting but unresolved question regards the genesis of the current sheets which ultimately reconnect. Given the strong link between the correlation scale turbulence increments and the reconnection magnetic fields, reconnecting current sheets in turbulence may often be occurring between correlation scale structures. In fact, in our simplified model for direct reconnection heating, we have effectively assumed that every
large-scale flux rope in the system is currently undergoing reconnection.
This assumption leads to nearly all of the plasma in the system interacting with a reconnection region.
Additionally, for systems with high (effective) Reynolds numbers,
many more small-scale or ``secondary'' flux ropes are likely also present
\cite{WanEA13-xpts,Haggerty2017ExploringTurbulence},
and these are not explicitly included in our estimations.
Reconnection heating models for turbulence have also been generalized~\cite{StawarzEA22} to include a prefactor equal to the fraction of particles that interact with a reconnection region.
  Recalling
that the inverse cascade of magnetic flux in 2D systems tends to lead to larger and larger flux ropes
~\citep{MattMont80}, this
  ``space-filling''
assumption is likely reasonable, but will require significant further investigation.

\begin{acknowledgments}
This research is supported by the NASA grants 80NSSC20K1813 (M.A.S.), 80NSSC24K0172 (M.A.S.) and 80NSSC24K0559 (M.A.S.); 80NSSC23K0409 (P.A.C.);
80NSSC21K1458 (R.B.), awarded under the MMS Early Career program; 80NSSC25K7757 (Y.Y.); 80NSSC23K0099 (J.M.T.); and the MMS Mission, Theory, Modeling and Data Analysis Program grant 80NSSC19K0565 (W.H.M.). It is also supported by NSF grant PHY-2108834 (W.H.M.); and 2347952 (R.B.), through the FDSS Track 1 award for the Expansion and Integration of Space Physics at the University of Delaware; 2501387 (F.P.), through the SHINE program at the University of Delaware; PHY-2607428 (P.A.C.); PHY-2607293 (C.C.H.); NSF-CAREER Grant AGS-2338131 (C.C.H.) and by the Royal
Society University Research Fellowship URF\textbackslash R\textbackslash 251029 (J.E.S.). We acknowledge high-performance computing support from the Derecho system \href{https://doi.org/10.5065/qx9a-pg09}{(https://doi.org/10.5065/qx9a-pg09)} provided by the NSF National Center for Atmospheric Research (NCAR), sponsored by the National Science Foundation. We also acknowledge the National Energy Research Scientific Computing Center (NERSC), a U.S. Department of Energy Office of Science User Facility operated under Contract No.~DE-AC02-05CH11231.
\end{acknowledgments}

\section*{Data availability}
MMS data are publicly available from the MMS Science Data Center
\url{https://lasp.colorado.edu/mms/sdc/public/}. The reconnection event list
analyzed here was published previously in ref.~\cite{StawarzEA22} and is
provided in that paper's supplementary material. PIC simulation data and
analysis scripts are available from the corresponding author upon
reasonable request.

\section*{Author contributions}
M.B.K., M.A.S., W.H.M., and R.B. designed research; M.B.K., M.A.S., and W.H.M. performed research; M.B.K. analyzed data; M.B.K., S.A., J.E.S., F.P., C.C.H., and Y.Y. contributed data and analysis methods; S.O., P.A.C., S.R., and J.M.T. reviewed and advised; and M.B.K., M.A.S., W.H.M., and S.O. wrote the paper.

\section*{Competing interests}
The authors declare no competing interest.

 \newcommand{\BIBand} {and} 
  \newcommand{\boldVol}[1] {\textbf{#1}} 
  \providecommand{\SortNoop}[1]{} 
  \providecommand{\sortnoop}[1]{} 
  \newcommand{\stereo} {\emph{{S}{T}{E}{R}{E}{O}}} 
  \newcommand{\au} {{A}{U}\ } 
  \newcommand{\AU} {{A}{U}\ } 
  \newcommand{\MHD} {{M}{H}{D}\ } 
  \newcommand{\mhd} {{M}{H}{D}\ } 
  \newcommand{\RMHD} {{R}{M}{H}{D}\ } 
  \newcommand{\rmhd} {{R}{M}{H}{D}\ } 
  \newcommand{\wkb} {{W}{K}{B}\ } 
  \newcommand{\alfven} {{A}lfv{\'e}n\ } 
  \newcommand{\alfvenic} {{A}lfv{\'e}nic\ } 
  \newcommand{\Alfven} {{A}lfv{\'e}n\ } 
  \newcommand{\Alfvenic} {{A}lfv{\'e}nic\ }
\end{document}